\documentclass[12pt,preprint]{aastex}

\newcommand{\maps}{Met. Planet. Sci.}
\newcommand{\ppiv}{Protostars and Planets IV}
\newcommand{\lpsc}{Lunar Planet. Sci. Conf.}
\newcommand{\phil}{Phil. Trans. Royal Soc. A}

\shorttitle{Early Solar System Irradiation} \shortauthors{Gounelle
et al.}

\begin{document}


\title{The Irradiation Origin of Beryllium Radioisotopes and Other Short-lived Radionuclides}

\author{Matthieu Gounelle\altaffilmark{1,2,3}, Frank H. Shu\altaffilmark{4},
Hsien Shang\altaffilmark{5}, A. E. Glassgold\altaffilmark{6}, K.
E. Rehm\altaffilmark{7}, Typhoon Lee\altaffilmark{8}}

\altaffiltext{1}{Universit\'e Paris XI-Centre de Spectrom\'etrie
Nucl\'eaire et de Spectrom\'etrie de Masse (CSNSM), B\^ atiment
104, 91 405 Orsay Campus, France }\altaffiltext{2}{Also at Impact
\& Astromaterials Research Center (IARC), Department of
Mineralogy, Natural History Museum, London,
UK}\altaffiltext{3}{Present address: Laboratoire d'\'{E}tude la
Mati\`ere Extraterrestre, Mus\'{e}um National d'Histoire
Naturelle, 57 rue Cuvier, 75 005 Paris} \altaffiltext{4}{National
Tsing Hua University, Hsinchu, Taiwan}
 \altaffiltext{5}{Institute of
Astronomy and Astrophysics, Academia Sinica, Taipei, Taiwan}
\altaffiltext{6}{Department of Astronomy, University of
California, Berkeley, Berkeley, CA 94720-3411, USA}
\altaffiltext{7}{Physics Division, Argonne National Laboratory
Argonne, IL 60439, USA} \altaffiltext{8}{Institute of Earth
Science, Academia Sinica, Taipei 115, Taiwan}

\begin{abstract}
Two explanations exist for the short-lived radionuclides
($T_{\frac{1}{2}}$ $\leq$ 5 Ma) present in the solar system when
the calcium-aluminum-rich inclusions (CAIs) first formed. They
originated either from the ejecta of a supernova or by the {\it in
situ} irradiation of nebular dust by energetic particles. With a
half-life of only 53 days, Beryllium-7 is then the key
discriminant, since it can  be made only by irradiation. Using the
same irradiation model developed earlier by our group, we
calculate the yield of $^7$Be. Within model uncertainties
associated mainly with nuclear cross sections, we obtain agreement
with the experimental value. Moreover, if $^7$Be and $^{10}$Be
have the same origin, the irradiation time must be short (a few
to tens of years), and the proton flux must be of order $F
\sim$ 2 $\times$ 10$^{10}$ cm$^{-2}$ s$^{-1}$. The x-wind model
provides a natural astrophysical setting that gives the requisite
conditions.  In the same irradiation environment, $^{26}$Al,
$^{36}$Cl and $^{53}$Mn are also generated at the measured levels
within model uncertainties, provided that irradiation occurs under
conditions reminiscent of solar impulsive events (steep energy
spectra and high $^3$He abundance). The decoupling of the
$^{26}$Al and $^{10}$Be observed in some rare CAIs receives a
quantitative explanation when rare gradual events (shallow energy
spectra and low $^3$He abundance) are considered. The yields of
$^{41}$Ca are compatible with an initial solar system value inferred
from the measured initial $^{41}$Ca/$^{40}$Ca ratio and an
estimate of the thermal metamorphism time (Young et al. 2005), alleviating
the need for two-layer protoCAIs. Finally, we show that the
presence of supernova-produced $^{60}$Fe in the solar accretion
disk does not necessarily mean that other short-lived
radionuclides have a stellar origin.
\end{abstract}


\newpage

\section{Introduction}
An important result concerning the formation of the solar system
is the discovery that radionuclides with half-lives of the order
of a million years (Ma) and less were alive when
Calcium-Aluminum-rich Inclusions (CAIs) and chondrules, the
igneous components of chondrites, were formed (see Russell,
Gounelle \& Hutchinson 2001 and McKeegan \& Davis 2003 for recent
reviews). Furthermore, some short-lived radioactivities, $^{10}$Be
($T_{\frac{1}{2}}$ = 1.5 Ma), $^{26}$Al (0.74 Ma), $^{36}$Cl (0.3
Ma), $^{41}$Ca (0.1 Ma), $^{60}$Fe (1.5 Ma) and possibly $^{53}$Mn
(3.7 Ma), existed at levels (see Table 1) well above the
expectations of models of continuous galactic nucleosynthesis
(e.g. Meyer \& Clayton 2000). These findings have motivated two
different kinds of explanations. The {\it external} model
stipulates that the radionuclides were made in a supernova or in
an asymptotic giant branch (AGB) star (e.g., Busso, Gallino \&
Wasserburg 2003). In the case of a supernova, it has been proposed
that the expelled material triggered the collapse of a nearby
molecular cloud core, which led to the formation of the solar
system (Cameron \& Truran 1977). More recently, an Orion-like
environment has been suggested for the birth of the solar system
(Hester et al. 2004). The {\it internal} irradiation model
conjectures that the radionuclides were created by the irradiation
of the solar environment by energetic protons, $^4$He and $^3$He
nuclei, accelerated by gradual or impulsive events near an active
young Sun (Lee et al. 1998; Goswami et al. 2001; Gounelle et al.
2001; Marhas et al. 2002; Leya et al. 2003). Thus, establishing
the origin of the short-lived radionuclides can provide important
constraints on the astrophysical setting of the Sun's birth,
stellar nucleosynthesis models, irradiation processes around young
stellar objects, and early solar system chronology (Gounelle et
al. 2005a,b; Kita et al. 2005).

Beryllium-7 decays to $^7$Li with a half-life of 53 days. Because
of this very short half-life, it must have an internal irradiation
origin, since any $^7$Be produced outside the nascent solar system
would have decayed long before being incorporated into the first
condensing solids. While the mere discovery of $^7$Be would be
definite proof for {\it in-situ} irradiation, measurement of its
abundance, together with that of other irradiation products such
as $^{10}$Be (McKeegan Chaussidon and Robert~2000), would provide
important information concerning the irradiation conditions
(duration as well as flux, energy spectra and isotopic composition
of the accelerated particles).

The experimental detection of $^7$Be has proven to be a long and
difficult quest, largely because of the high mobility of its
daughter element, Li. A hint of the presence of $^7$Be was first
seen in an Allende CAI (USNM 3515) at a high relative abundance,
$^7$Be/$^9$Be $\sim$ 0.1 (Chaussidon, Robert, \& McKeegan 2002).
However, a definitive isochron could not be established for this
CAI, which has been highly disturbed by secondary events.
Recently, Chaussidon, Robert \& McKeegan (2004, 2005) identified
37 spots within Allende CAI 3529-41 for which the Li isotopic
record has been undisturbed since crystallization. They determined
a well-defined isochron for these 37 points and established that
$^7$Be was alive in the early solar system at a level
$^7$Be/$^9$Be = $(6.1 \pm 1.3)\times 10^{-3}$ (2$\sigma$ error).

The goal of this paper is to examine the yields of $^7$Be in the
context of the irradiation model we have developed over the past
several years (Shu et al. 1996; Shu et al. 1997; Lee et al. 1998;
Gounelle et al. 2001). We will establish additional constraints on
irradiation models imposed by the presence of $^7$Be and $^{10}$Be in
the early solar system. Calculations for the production of $^{36}$Cl,
a newly discovered radionuclide (Lin et al. 2005) with a half-life of
0.3 Ma, are also presented, as well as new data and updated
computations for $^{26}$Al, $^{41}$Ca and $^{53}$Mn.  We also discuss
the implications of the high abundance of $^{60}$Fe in the early solar
system.

\section{The Irradiation Model}
\label{sec-model}

\subsection{Outline of the Model}

We use the irradiation model in the form developed by Gounelle et
al.~(2001) to calculate the yield of $^7$Be. Their preferred case
(c.f. Figure 2d of that paper) is adopted, and the parameters are
kept the same except for the spectral index $p$ and the
$^3$He/$^1$H ratio, which are varied within reasonable ranges.
Case 2d refers to the geometry of the assumed core-mantle
structure of protoCAIs: the minimum and maximum core sizes vary
between 0.0050 and 2.5 cm, as natural CAIs do, and the mantle
thickness is fixed at 0.28\,cm (Shu et al.~2001). Here we provide
the main characteristics of our model and discuss how it is
implemented for the new calculations of $^7$Be and $^{36}$Cl. The
very short half-life of $^7$Be (53 d vs $\sim$ Ma for other
radionuclides) calls for a special treatment. Additional details
on the irradiation model may be found in Gounelle et al.~(2001).

Of basic importance in irradiation models are the {\it absolute}
yields of the short-lived radionuclides. By absolute yield we mean the
abundance of a radionuclide calculated relative to a fixed,
stable-isotope reference (e.g. $^{26}$Al/$^{27}$Al), {\it not}
abundances of radionuclides relative to one another. The absolute
yields depend linearly on the product of the number flux of cosmic
rays, $F_{\rm CR}$, and the irradiation time, $\Delta t$. These quantities
can vary over many orders of magnitude, depending on the astrophysical
environment, implying changes in the short-lived-radionuclide yields
of many orders of magnitude. Other parameters, like the composite
structure of the targets, the shielding, or the cosmic ray properties,
induce variations in the yields of the short-lived radionuclides that
are no larger than a few orders of magnitude (Lee et al. 1998; Goswami
et al. 2001; Gounelle et al. 2001; Marhas et al. 2002; Leya et
al. 2003).

In our model, the number flux of protons is scaled to X-ray
observations of young stellar objects (Lee et al.~1998), with
$F_{\rm p}$ ($E \geq$ 10 MeV) $\sim$ 1.9 $\times$ 10$^{10}$
cm$^2$~s$^{-1}$. The irradiation of proto-CAIs occurs within the
reconnection ring (Lee et al. 1998), and the effective number of
proton passages in the reconnection ring is $\cal{M}$ = 1.6.
The irradiation time depends linearly on the size of the
proto-CAIs. With a typical irradiation time of 20 yr for a 1 cm
protoCAI, the irradiation time varies between a few years and a
few tens of years, since protoCAIs vary between 0.2805 and 2.78 cm
(Gounelle et al. 2001; Shu et al. 2001). Homogenization of the
irradiation products is assumed to take place through frequent
evaporation-condensation episodes  that, in a steady-state
picture, erase the calculated differences of the yields between
small and large CAIs. The final yield of a short-lived
radionuclide is obtained by integrating the irradiation products
over the size distribution of the total population of CAIs,
reflecting the continuous homogenization of irradiation products
made possible by the high frequency of large flares compared to
the irradiation duration (Grosso et al. 1998; Wolk et al. 2005).

The differential distribution of protons, $N(E)$, is assumed to be
a power-law in energy. The spectral index $p$ in $N(E) \propto
E^{-p}$ has a range between 2.7 and 5.  The number of $^4$He
nuclei relative to protons is fixed at 0.1, but the $^3$He/$^1$H
ratio varies between 0 and 1. Impulsive events are characterized
by relatively steep proton spectra (higher $p$) and large
abundances of $^3$He, while gradual events have shallower proton
spectra (lower $p$) and smaller abundances of $^3$He (Reames
1995). We envision two different structures for proto-CAIs, a
core-mantle configuration and a homogeneous chondritic composition
(Gounelle et al.~2001).

For the production of $^{36}$Cl, we consider proton, alpha and
$^3$He induced reactions on a diversity of targets such as Cl, S
and K. The adopted average chondritic abundance of Cl is 704 ppm
(Lodders 2003) while the adopted median CAI abundance is 390 ppm
(Sylvester et al. 1993).

For the production of $^7$Be (and $^{10}$Be) we consider proton,
alpha and $^3$He induced reactions on $^{16}$O. Because its
mean-life $\tau$ (76 days) is short compared to typical
irradiation times $\Delta t$, ranging from a few years to a
few tens of years, $^7$Be yields need to be calculated by
balancing spontaneous decay against production by spallation
reactions. Thus, for $^7$Be, and any short-lived radionuclide for
which $\tau \ll \Delta t$, the yield is the production rate times
$\tau$ rather than times $\Delta t$, which applies when $\tau \gg
\Delta t$.

\subsection{Updated Cross Sections}

\begin{figure}
\plotone{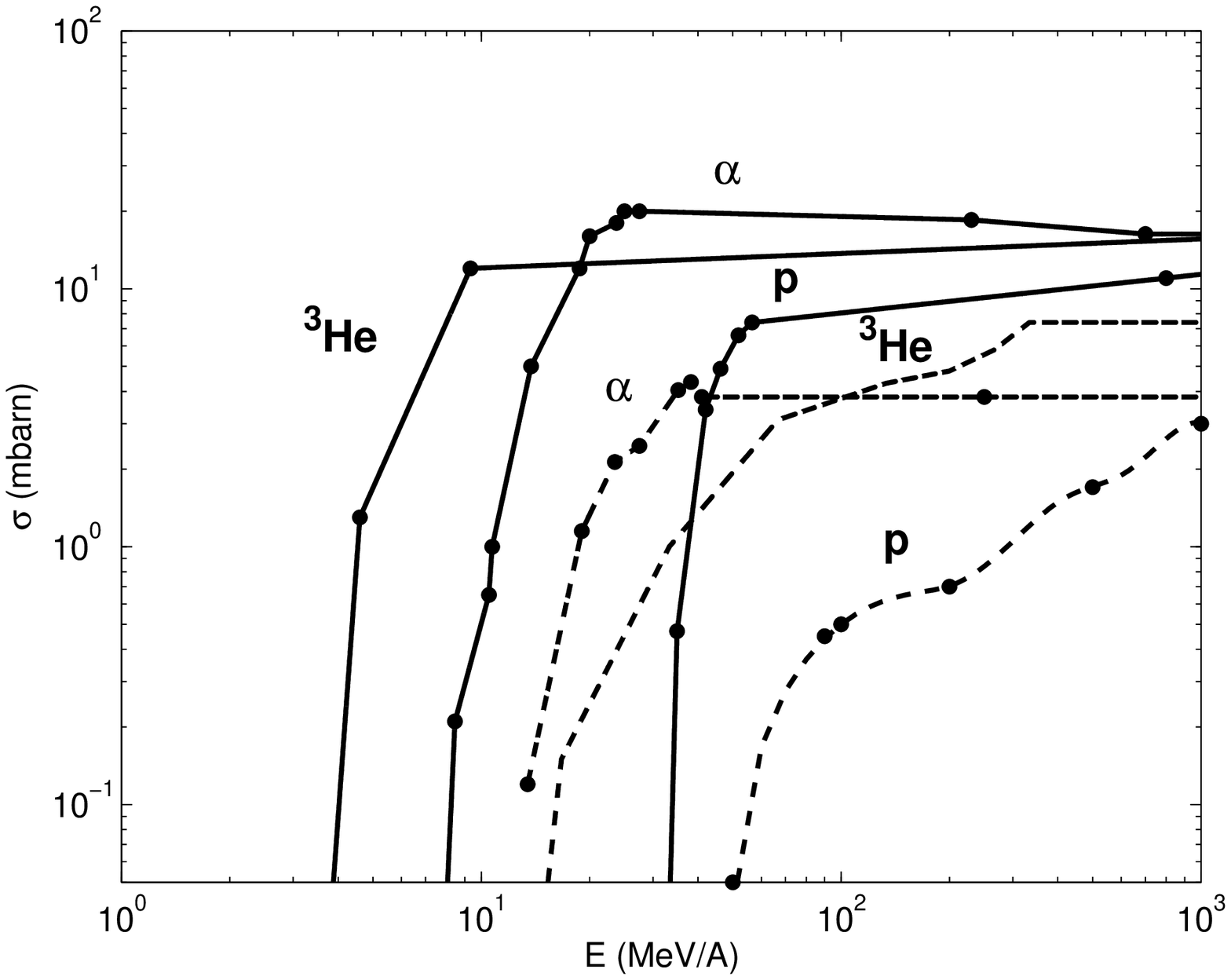} \caption{Adopted nuclear cross sections
for the reactions $^{16}$O(CR,x)$^7$Be and
$^{16}$O(CR,x)$^{10}$Be. The solid curves are the $^7$Be cross
sections, and the dashed lines are the $^{10}$Be cross sections.
The incident projectiles, CR = p, $^3$He, and $\alpha$ are
labelled on the graph. The nuclear cross sections are a
combination of experimental data and Hauser-Feshbach fragmentation
codes. Experimental data points are indicated by full black
circles. \label{fig-cross-be}}
\end{figure}

\begin{figure}
\plotone{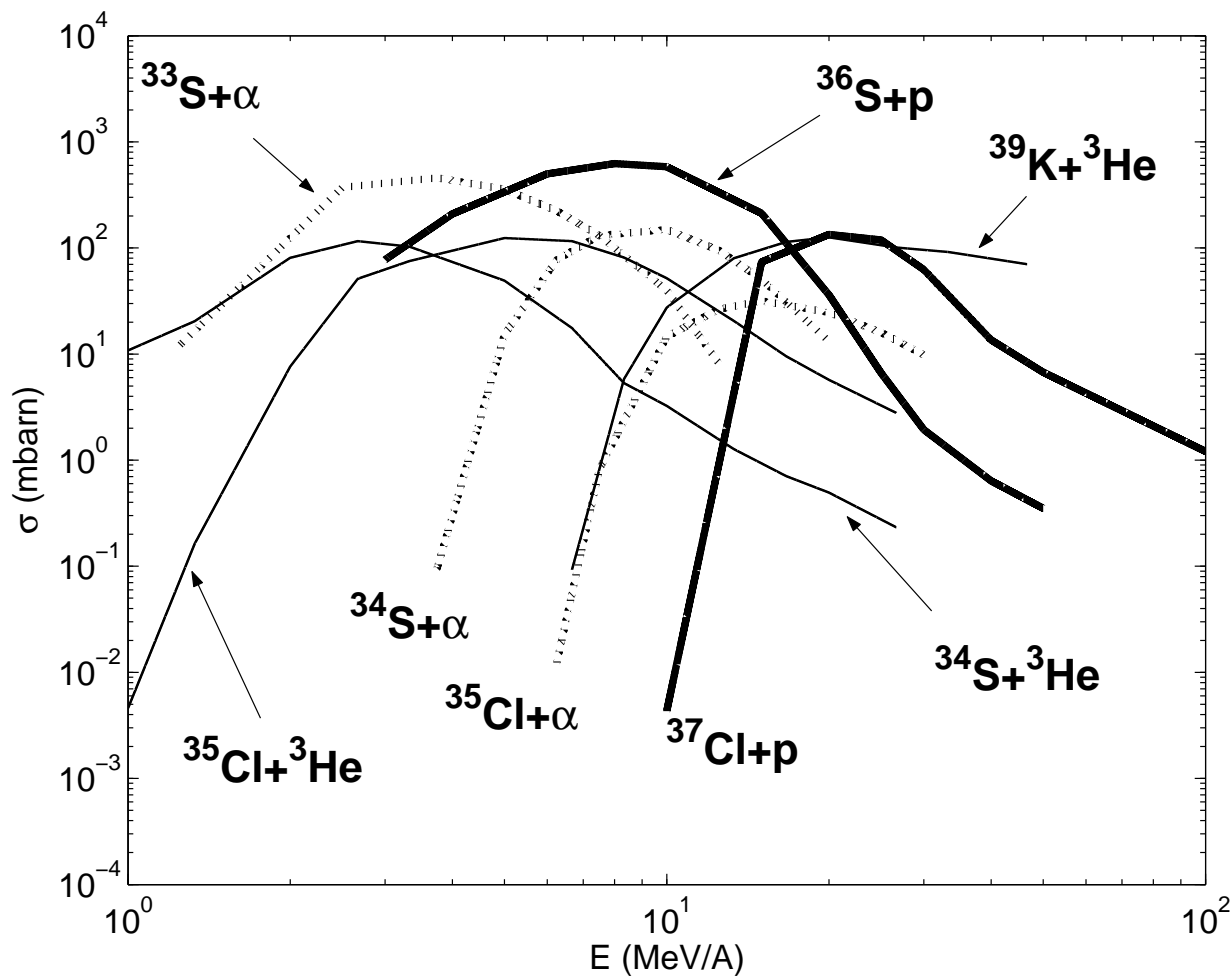} \caption{Adopted nuclear cross sections
for the proton (thick line), $\alpha$ (dotted line) and $^3$He
(dashed line) induced reactions for the production of $^{36}$Cl.
The cross sections are based on numerical simulations using
fragmentation and Hauser-Feshbach codes. In the absence of
experimental data points, the input parameters have been chosen so
that the calculations reproduce data obtained for lower and higher
masses (e.g. Al. Ca, Fe) where some data
exist.\label{fig-cross-cl}}
\end{figure}

The main new input data are the production cross sections shown in
Figure \ref{fig-cross-be} and Figure \ref{fig-cross-cl}. The
$^7$Be cross sections are based on numerical simulations using
fragmentation and Hauser-Feshbach codes and on experimental data
points (indicated by circles) from Landolt \& B\"ornstein (1994).
The $^{10}$Be-production cross-sections have been updated since
the work of Gounelle et al.~(2001). The $^{16}$O(p,x)$^{10}$Be
cross section is from Sisterson et al.~(1997); the
$^{16}$O($^4$He,x)$^{10}$Be cross section is from the experimental
study of Lange et al.~(1995) that had been previously overlooked;
and the $^{16}$O($^3$He,x)$^{10}$Be cross section is from a new
simulation.

The $^{36}$Cl-producing cross sections (Figure \ref{fig-cross-cl})
are based on numerical simulations using fragmentation and
Hauser-Feshbach codes. In the absence of experimental data,
the input model parameters have been chosen so that the calculations
reproduce data obtained for lower and higher masses (e.g. Al. Ca,
Fe) where some data exist. Because of the dependence of the cross
sections on the structure of the nuclei involved, it is difficult
to give precise uncertainties for these cross sections. From
comparisons with experimental data for other reactions, we estimate
an uncertainty of 50\% for the maximum of the cross section and a factor
of 2 for cross sections that are changing significantly with energy.

Most of the $^3$He cross-sections used by Lee et al.~(1998) and by
Gounelle et al.~(2001) for the production of short-lived
radionuclides came from theoretical nuclear physics codes and not
from direct laboratory measurements. To help reduce the
uncertainties based on this over-reliance on theory, Fitoussi et
al.~(2004) measured the cross section for the reaction
$^{24}$Mg($^3$He,p)$^{26}$Al that is the main source of $^{26}$Al.
In addition to using this experimentally determined cross section,
we also include proton and alpha reactions for the production of
$^{26}$Al and $^{41}$Ca that were previously ignored (for
impulsive events) on the basis that their contributions to the
$^{26}$Al and $^{41}$Ca yields were small compared with $^3$He.
These cross sections are: $^{26}$Mg(p,n)$^{26}$Al,
$^{27}$Al(p,pn)$^{26}$Al, $^{28}$Si(p,2pn)$^{26}$Al,
$^{24}$Mg($\alpha$,pn)$^{26}$Al, $^{42}$Ca(p,pn)$^{41}$Ca,
$^{40}$Ca($\alpha$,$^3$He)$^{41}$Ca,
$^{40}$Ca($\alpha$,$^3$H)$^{41}$Sc, and they can be found in Lee
et al.~(1998).

To summarize, we now use an improved and a more complete set of
cross sections compared to Gounelle et al.~(2001). Most of the key
cross sections for the production of $^7$Be, $^{10}$Be and
$^{26}$Al are now experimentally constrained. Among the cross
sections important for our calculations, only
$^{16}$O($^3$He,x)$^{10}$Be, $^{40}$Ca($^3$He,pn)$^{41}$Ca as well
as the $^{36}$Cl producing cross sections lack experimental data.
The uncertainties in the cross sections in Figure
\ref{fig-cross-be} (where experimental points are labeled with
filled circles) are a factor of the order of a few, and possibly
much larger for cross sections without experimental measurements
such as the ones presented in Figure \ref{fig-cross-cl}.

\subsection{Results of the Calculations}

\begin{figure}
\plotone{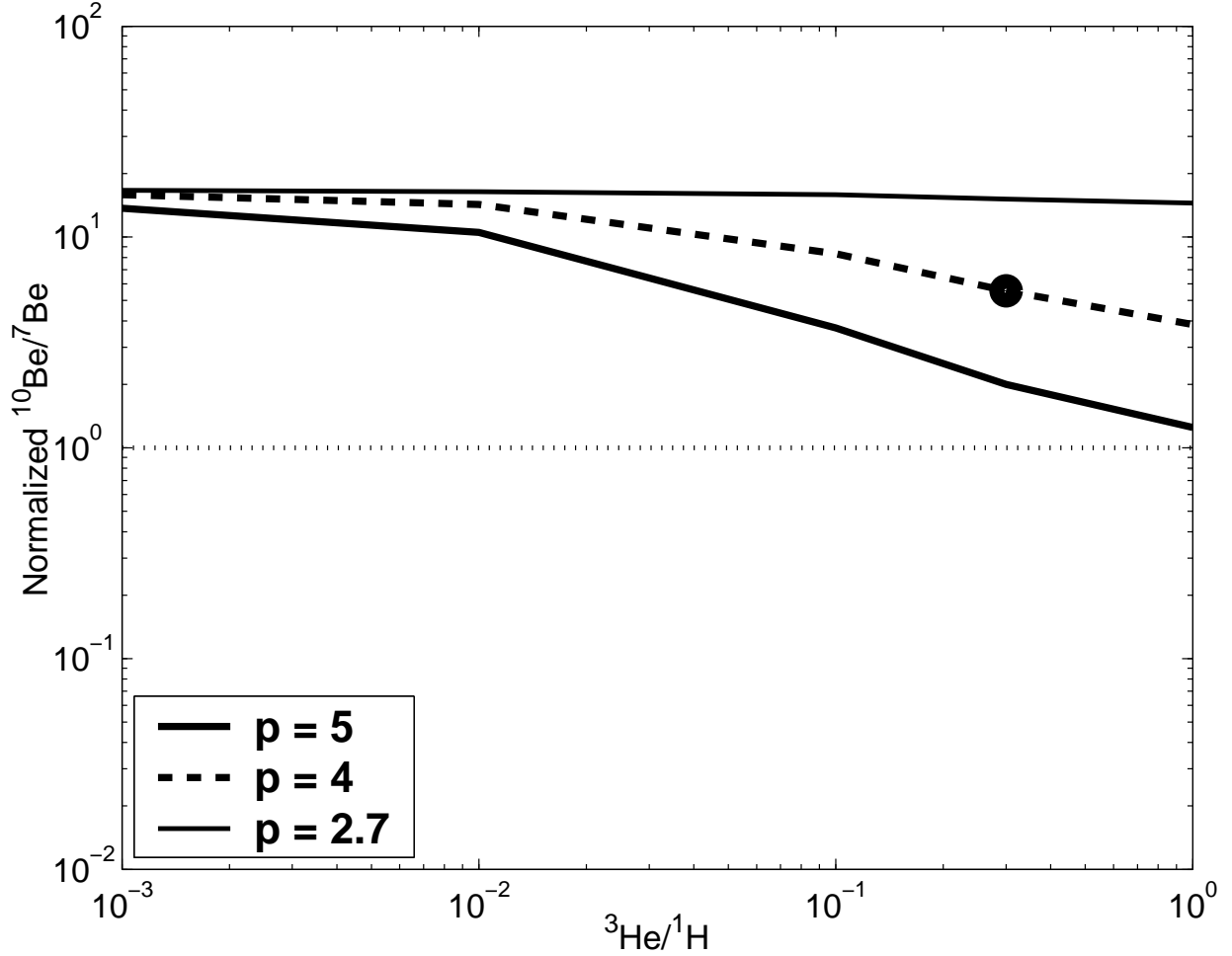} \caption{Plot of the $^{10}$Be/$^7$Be
ratio, normalized to the measured solar system value, as a
function of the spectral parameter $p$ and the $^3$He/$^1$H
abundance ratio. The
horizontal dotted line denotes the experimental value calculated
from Chaussidon, Robert and McKeegan (2005) and McKeegan,
Chaussidon and Robert (2001). The filled circle corresponds to the
preferred model of Gounelle et al.~(2001), i.e  with p = 4 and
$^3$He/$^1$H = 0.3. As in the case of $^7$Be (Table 2), the
dependence on the target composition is weak and is not shown
here. \label{fig-be-be}}
\end{figure}

Table 2 gives the calculated {\it{absolute}} yields of $^7$Be as a
function of the spectral index $p$ and the $^3$He/$^1$H ratio for
two different cases: a core-mantle and a homogeneous
chondritic compositional structure. The yields have been normalized to the
experimental value obtained by Chaussidon, Robert and McKeegan
(2005).  Figure \ref{fig-be-be} shows the (relative)
$^{10}$Be/$^{7}$Be ratio as a function of the $^3$He/$^1$H ratio
for a range of values of the spectral index $p$. Using the
{\it{absolute}} $^7$Be yields in Table 2, this is equivalent to
giving the {\it{absolute}} yield of $^{10}$Be. The result obtained
with the same parameters as Gounelle et al. (2001), corresponding
to $^3$He/$^1$H = 0.3 and $p$ = 4, is marked by a filled circle in
Figure \ref{fig-be-be}.

\begin{figure}
\plotone{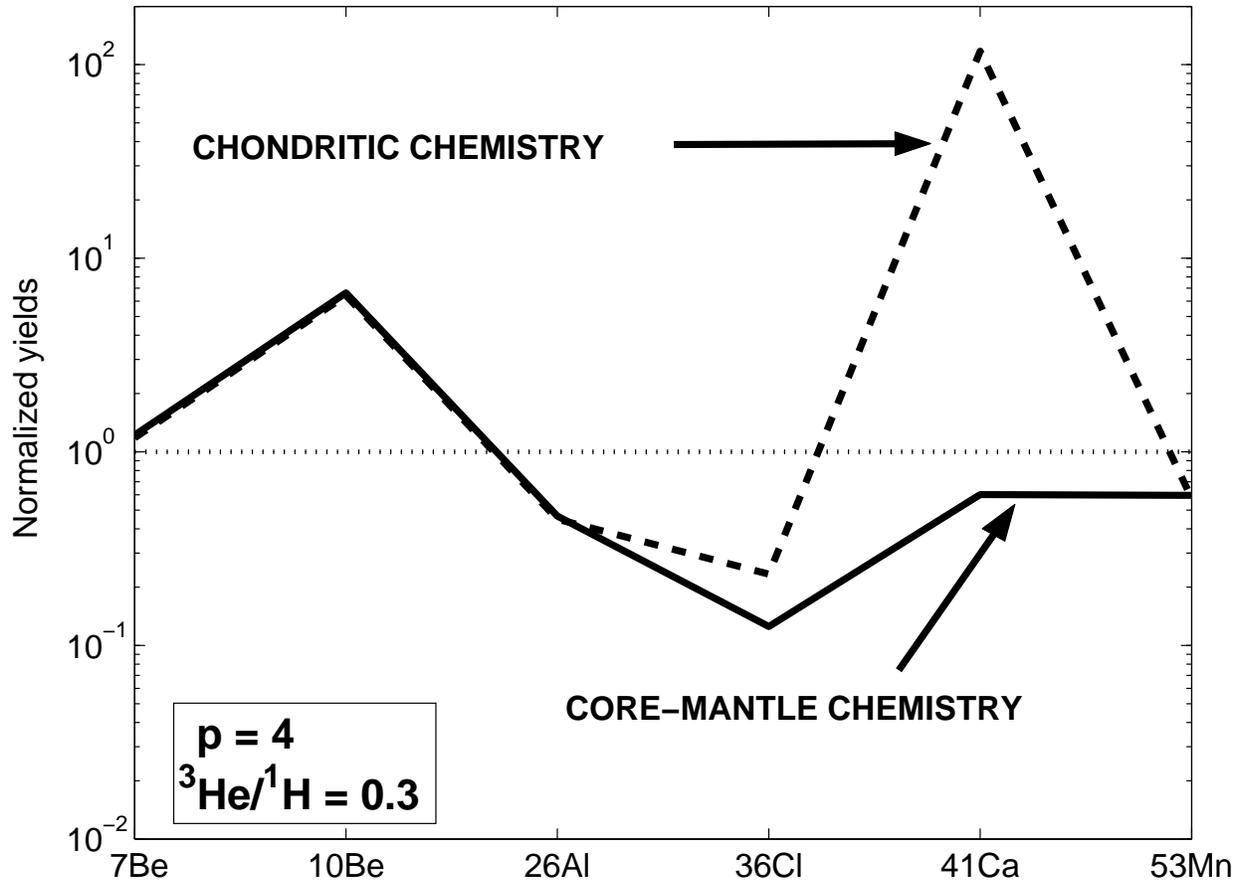} \caption{Yields of $^7$Be, $^{10}$Be,
$^{26}$Al, $^{36}$Cl, $^{41}$Ca and $^{53}$Mn for fixed spectral
parameters (p = 4, $^3$He/$^1$H = 0.3) and two different chemical
compositions (core-mantle and chondritic). The yields are
normalized to the experimental values (see Table 1), corresponding
to the horizontal dotted line. \label{fig-all2}}
\end{figure}

Figure \ref{fig-all2} shows the irradiation yields for the
short-lived radionuclides of primary interest, $^7$Be, $^{10}$Be,
$^{26}$Al, $^{36}$Cl,$^{41}$Ca, $^{53}$Mn, for the preferred
parameters, $p = 4$ and $^3$He/$^1$H = 0.3, and for the two
different target compositions (core-mantle and chondritic). Figure
\ref{fig-all3} presents the results for a variety of spectral
parameters. The case $p = 2.7$ and $^3$He/$^1$H = 0 represents
typical gradual flares, while the case $p = 4$ and $p = 5$ with
$^3$He/$^1$H = 0.3 represents impulsive flares.  We now discuss
the implications of these results.

\section{Discussion}
\label{sec-discussion}

\subsection{The Irradiation Origin of $^7$Be}

The detection of $^7$Be, with a half-life of 53\,d, in meteorites
(Chaussidon, Robert and McKeegan~2005) provides definitive evidence
for {\it in situ} irradiation of pre-meteoritic solids and for
local irradiation models. In fact, an earlier theoretical
calculation, using preliminary cross section data (Gounelle et
al.~2003), {\it predicted} the experimental result (Chaussidon,
Robert \& McKeegan~2004) to within a factor of two.
An explanation of $^7$Be in terms of
some other process (e.g., Cameron 2003) is possible only if CAIs themselves
formed outside the solar system and were never subsequently re-heated to reset the
beryllium-lithium clock. In such a case, however,
the entire dating of the solar system, based on the nearly
identical radiochemical ages of all CAIs, would be called into
question.

For reasonable cosmic ray parameters ($p$ ranging between 2.7 and
5, and $^3$He/$^1$H ranging from 0 to 1), the $^7$Be yields in
Table 2 vary between 0.12 and 5.6 times the experimental value.
The preferred model of Gounelle et al. (2001), with $p = 4$ and
$^3$He/$^1$H = 0.3, gives $^7$Be/$^9$Be = 7.5 $\times$ 10$^{-3}$,
which compares favorably with the measured value of $^7$Be/$^9$Be
= 6.1 $\times$ 10$^{-3}$. As can be seen in Table 2, the $^7$Be
yields do not depend very much on the chemical composition of the
target (usually no more than a few per cent). This is because they depend
on the O/Be ratio, and refractory Be undergoes minimal chemical
fractionation. Although the agreement between the model and the
experimental value is very good, uncertainties in the model,
especially in the nuclear cross sections (see section
\ref{sec-model}), call for caution.

A possible source for variations in the yields of $^7$Be are
fluctuations of the energetic particle flux. Because the relevant
timescale for $^7$Be synthesis is tens of days, $^7$Be is more
likely to be sensitive to such fluctuations than, say, $^{10}$Be
with a half-life of 1.5Ma. Significant X-ray flares are observed
for Sun-like young stellar objects with ages in the 1-2\,Ma range
on the order of several days (Wolk et al.~2005). Thus $^7$Be
synthesis in natural samples is likely to be more variable than other short-lived
radionuclides.

A basic reason for the good agreement between the present model
calculations and experiment is that the production of $^7$Be
requires a high accelerated particle flux. As pointed out in \S
\ref{sec-model}), the yield of $^7$Be with its short half-life
does not depend on the irradiation time $\Delta t$ but on the mean
life $\tau$. It also depends only moderately on the cosmic-ray
parameters, as can be seen in Table 2. Consistent with the results
in Table 2, the only way to produce $^7$Be at a significant level
is to have a high flux of accelerated particles. The value adopted
by Gounelle et al.~(2001) and used here, $F_{\rm p}(E \geq E_{10})
\sim 1.9 \times 10^{10}$ cm$^2$~s$^{-1}$, is consistent with the
measured X-ray properties of Sun-like stars (e.g., Wolk et
al~2005). Our local irradiation scenario is also naturally conducive
to high accelerated-particle fluxes because the particles are confined
to as well as produced within the reconnection ring (the irradiation
zone) by strong magnetic fields (see \ref{sec-context}).

\subsection{The Case of $^{10}$Be}

\begin{figure}

\plotone{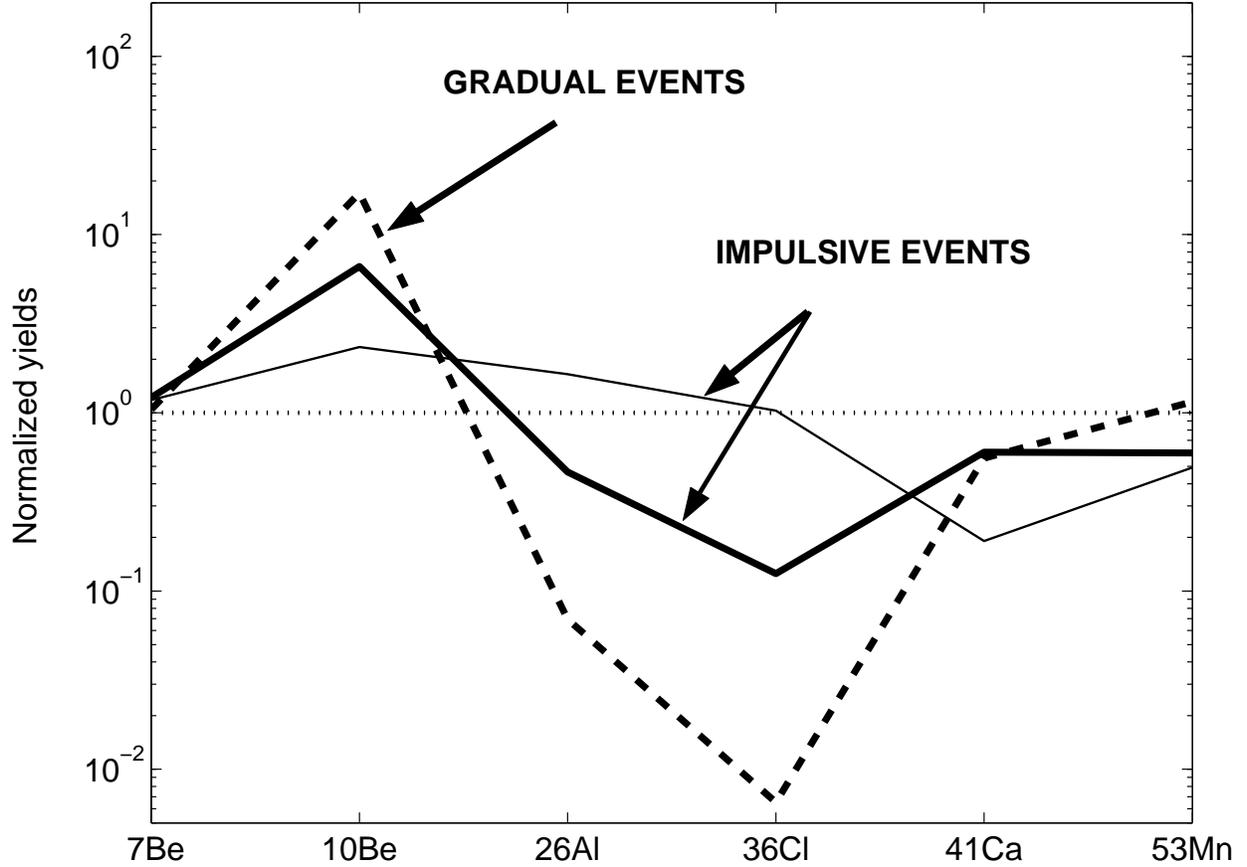} \caption{Yields of $^7$Be, $^{10}$Be,
$^{26}$Al, $^{36}$Cl, $^{41}$Ca and $^{53}$Mn for fixed
core-mantle structure (case 2d of Gounelle et al. (2001)) and a
variety of spectral parameters. Impulsive flares have $p = 4$
(thick line) or $p=5$ (thin line), and $^3$He/$^1$H = 0.3 while
gradual flares have $p = 2.7$ and $^3$He/$^1$H = 0 (dashed line).
The yields are normalized to the experimental values (see Table
1), corresponding to the horizontal dotted line.\label{fig-all3}}
\end{figure}

Our results for $^{10}$Be are shown in Figures \ref{fig-be-be},
\ref{fig-all2} and \ref{fig-all3}. The calculated
$^{10}$Be/$^{7}$Be ratio depends moderately strongly on the
cosmic-ray parameters (index $p$ and $^3$He/$^1$H), when
$^3$He/$^1$H $>0.01$, as shown in
Figure \ref{fig-be-be}. The preferred model of Gounelle et
al.~(2001) gives a satisfactory account of the experimental ratio:
($^{10}$Be/$^{7}$Be)$_{\rm model}$ = 0.88 vs.
($^{10}$Be/$^{7}$Be)$_{\rm exp}$ = 0.16. The over-production of
$^{10}$Be by a factor of $\sim$ 5 is not that large, given the
uncertainties in the cross sections. Moreover, it is reduced if we
use a somewhat steeper spectral index, e.g., $p=5$ as shown in
Figure \ref{fig-all3}. Alternatively, the $^{10}$Be yield would be
lower if the irradiation time was lower. However, this
modification would also decrease the yields of other short-lived
radionuclides such as $^{26}$Al.

The ratio $^{10}$Be/$^{7}$Be does not depend at all on the number
flux (see section \S 2.1) and very little on the target
composition (see Table 2). It depends mostly on the irradiation
time, which is estimated in our 2001 model to be a few years to a
few tens of years. The irradiation time is essentially the
residence time for proto-CAIs of a certain size in the
reconnection ring before ejection to asteroidal distances by the
fluctuating X-wind (Shu et al.~2001).  Such time scales are also
suggested by observations of the intervals between knot-forming
episodes in the optical jets of young stellar objects (Reipurth
and Bally 2001). We cannot emphasize too strongly that the
resulting {\it mechanical} model (Shu et al. 2001) yields in
order-of-magnitude exactly the irradiation time to reproduce the
$^{10}$Be/$^{7}$Be ratio.

Desch, Connolly and Srinivasan (2004) have challenged the idea
that $^{10}$Be was formed in the solar system by irradiation. They
argue instead that most of the solar system $^{10}$Be originated
as $^{10}$Be nuclei in galactic cosmic rays stopped in the
progenitor molecular cloud core. This conclusion depends
sensitively on a number of assumptions made in their calculations,
e.g., in the variation of the mass-column density
$\Sigma(\varpi,t)$ in a collapsing molecular-cloud core, taken
from an earlier numerical study by Desch \& Mouschovias (2001).
Desch et al. (2004) regard $\Sigma$ as being spatially uniform,
whereas Desch \& Mouschovias (2001) found $\Sigma$ to decrease
considerably with increasing distance $\varpi$ from the center of
the cloud core. The core has a mass (45 $\rm M_{\odot}$),
significantly in excess of what is expected for the formation of a
solar-mass star. Furthermore, the time scale for core formation
and $^{10}$Be accumulation is $\sim 10$\,Ma, an order of magnitude
longer than allowed by the statistics of nearby molecular cloud
cores with and without embedded stars (e.g., Lee \& Myers 1999).
Even so, Desch et al. (2004) still need to increase the galactic
cosmic ray flux relevant to present values by an {\it ad hoc}
factor of 2 to produce enough $^{10}$Be. The long evolutionary
time also leaves no room for the injection into the nascent
protosolar cloud of other short-lived radionuclides like
$^{26}$Al, $^{41}$Ca and $^{53}$Mn from a stellar event.
Because of its ad hoc nature, the proposal of Desch, Connolly and
Srinivasan (2004) has few advantages over the more common view
that $^{10}$Be has an in-situ irradiation origin (e.g., McKeegan,
Chaussidon \& Robert~2000; Gounelle et al.~2001; Marhas et
al.~2002; Huss 2004).

\subsection{The Co-production of $^{26}$Al, $^{36}$Cl, $^{41}$Ca and $^{53}$Mn}

Our preferred model (case 2d of Gounelle et al.~2001) with $p = 4$
and $^3$He/$^1$H = 0.3) can reproduce within a factor of a few the
measured solar system abundance of $^7$Be, $^{10}$Be, $^{26}$Al,
$^{36}$Cl, $^{41}$Ca and $^{53}$Mn (the thick solid curve in Figure 5).
Chlorine-36 is slightly under produced relative to its experimental
value but, given the uncertainties of the model, the agreement is
satisfactory. The adoption of a new {\it experimental} cross section
for the reaction $^{24}$Mg($^3$He,p)$^{26}$Al does not modify the yield of
$^{26}$Al by more than a factor of 2, keeping the production of
this radio-isotope in line with what had been calculated using {\it
theoretical} cross sections. Taking into account the previously
ignored proton and $^4$He induced cross sections for impulsive
flares does not change the yields by more than ten percent.
However, the adoption of an experimental cross section for
$^{16}$O($^4$He,x)$^{10}$Be has slightly increased the $^{10}$Be
yield, degrading the quality of the fit with the experimental
data. Adoption of a somewhat steeper spectral index ($p = 5$
instead of $p = 4$, all other parameters remaining the same) would
put $^{10}$Be and $^{36}$Cl in line with the measured solar system
value. This would not change much the yields of the other
short-lived radionuclides (Figure \ref{fig-all3}), except for
$^{41}$Ca discussed below.

The yields of all of the main short-lived radionuclides (including
the newly calculated chlorine-36) are independent of composition
except for $^{41}$Ca, as may be seen in Figure \ref{fig-all2}. The
low $^{41}$Ca measured value, $^{41}$Ca/$^{40}$Ca = 1.5 $\times$
10$^{-8}$ (Srinivasan et al.~1994), can be reproduced by our model
only when a core-mantle chemistry is adopted (Shu et al.~1997;
Gounelle et al. 2001). However, there are hints that the solar
system initial $^{41}$Ca/$^{40}$Ca may have been considerably
higher (McKeegan et al.~2004) than the presently measured value.

Recent experimental findings suggest that the initial
$^{26}$Al/$^{27}$Al ratio of the solar system was higher than the
measured "canonical" value 4.5 $\times$ 10$^{-5}$ (MacPherson et
al. 1995), with an initial $^{26}$Al/$^{27}$Al of the order $\sim$
7 $\times$ 10$^{-5}$ (Young et al.~2002; Galy et al.~2004; Young
et al.~2004). Subsequent redistribution of Mg isotopes within the
CAI would then lead to an apparent initial ratio of 4.5 $\times$
10$^{-5}$, i.e., the canonical value. Given the uncertainties of
our irradiation model, an increase of $\sim$ 20\,\% in the initial
$^{26}$Al/$^{27}$Al ratio would not change the overall level of
agreement between the calculations and the measurements. However,
it might partly solve the overproduction of $^{41}$Ca observed
when a pure chondritic composition is adopted (Figure
\ref{fig-all3}).

\begin{figure}

\plotone{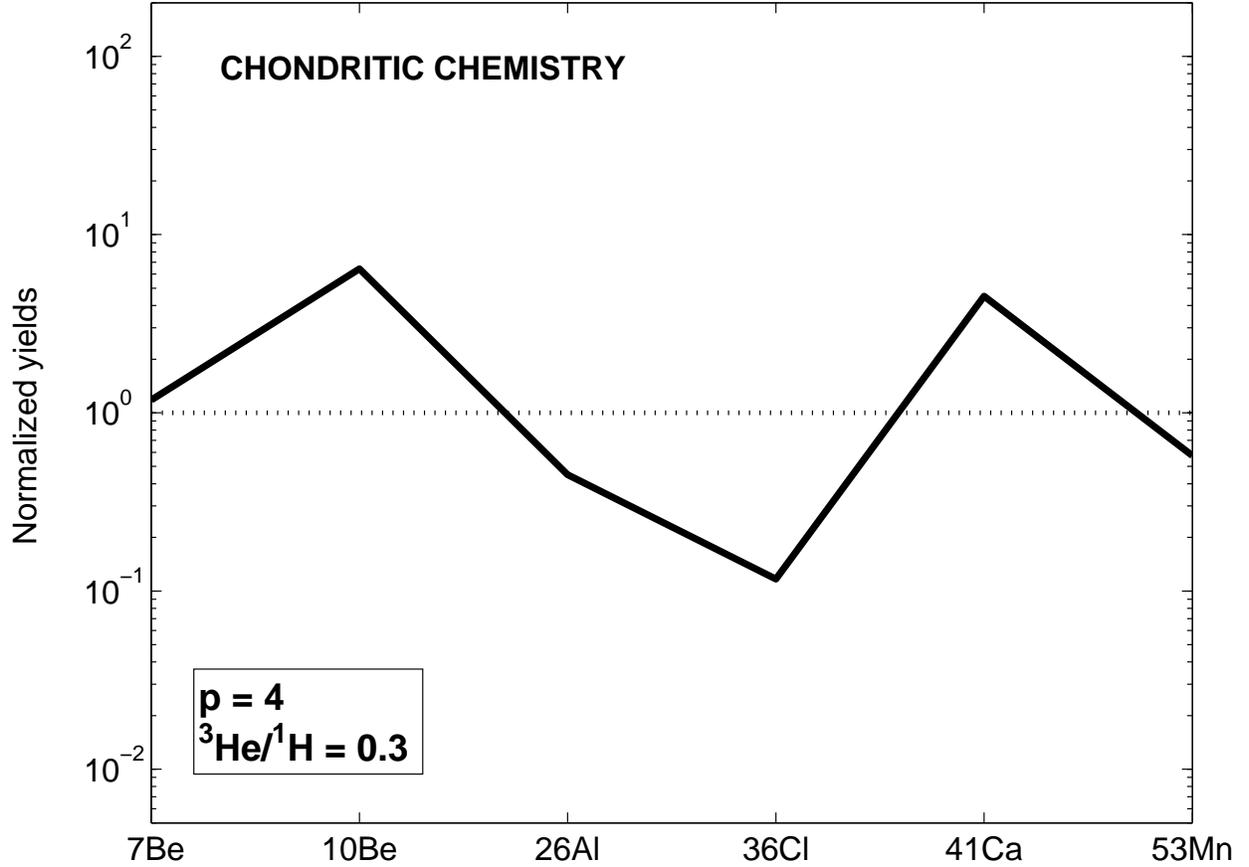} \caption{Yields of $^7$Be, $^{10}$Be,
$^{26}$Al, $^{36}$Cl, $^{41}$Ca and $^{53}$Mn for a chondritic
chemistry and the spectral parameters $p=4$ and $^3$He/$^1$H =
0.3. The yields are normalized to the experimental values (see
Table 1), corresponding to the horizontal dotted line. For
$^{41}$Ca, we have adopted the initial value $^{41}$Ca/$^{40}$Ca =
3.9 $\times$ 10$^{-7}$, more likely than the canonical ratio
depicted in Table 1 (see text). \label{fig-ca}}
\end{figure}

Calcium-41 decays into $^{41}$K (see Table 2). If the K isotopes
had been subjected to the same isotopic resetting as the Mg
isotopes, this would mean that the initial $^{41}$Ca/$^{40}$Ca
ratio might have been higher by a factor of ($^{26}$Al$_{\rm
initial}$/$^{26}$Al$_{\rm reset}$) $^{\frac{\tau (26Al)}{\tau
(41Ca)}}$ = (7/4.5) $^{\frac{\tau (26Al)}{\tau (41Ca)}}$ = 26,
where we have simply applied exponential decay law to both
$^{26}$Al and $^{41}$Ca. It is therefore very likely that the
initial $^{41}$Ca/$^{40}$Ca ratio was closer to 3.9 $\times$
10$^{-7}$ than to 1.5 $\times$ 10$^{-8}$. Such an early
solar-system value would be consistent, within errors, with our
computed irradiation yield of $^{41}$Ca without the ad hoc
adoption of a core-mantle structure (see Figure \ref{fig-ca}).
Isotopic redistribution would be less important for other
short-lived radionuclides because they have longer half-lives than
$^{26}$Al.  Chlorine-36 could also have been higher by a factor of
$\sim$ 3.8 because of this thermal metamorphism, slightly
degrading the good fit of Figures 5 and 6. Chaussidon, Robert and
McKeegan (2005) have observed such a post-magmatic redistribution
for Li isotopes, but have been able to deconvolve its effect from
the $^7$Be decay.

\subsection{Decoupling of $^{26}$Al and $^{10}$Be}

Mahras, Goswami \& Davis (2002) have detected the past presence of
$^{10}$Be in some rare refractory inclusions containing calcium
aluminum oxides known as hibonites in which $^{26}$Al and
$^{41}$Ca (Sahjipal et al.~1998) are conspicuously absent. They interpret this
``decoupling'' between $^{10}$Be and $^{26}$Al (and $^{41}$Ca)
as contradicting irradiation models that
simultaneously produce $^{10}$Be, $^{26}$Al and $^{41}$Ca.

Hibonites from CM2 chondrites, which show this decoupling
between $^{10}$Be and $^{26}$Al, share with FUN inclusions the
property of bearing large isotopic anomalies in neutron-rich
isotopes.  Such FUN inclusions also formed with little initial
$^{26}$Al, a fact accounted for by Lee et al.~(1998) who suggested
that FUN CAIs never experienced the impulsive-flare events that
are rich in $^3$He before being ejected by the X-wind.  However,
CAIs that never enter the reconnection ring can still be exposed
to gradual-flare events that occur at larger distances.  Such gradual
events are capable of producing large amounts of $^{10}$Be without
co-producing $^{26}$Al (or $^{41}$Ca), according to the results
shown in Figure \ref{fig-all3}.  Thus, the occurrence of rare CAIs
containing abundant $^{10}$Be but little $^{26}$Al or $^{41}$Ca
may have an explanation in exposure to gradual but not impulsive
events.We note that the anomalies in stable isotopes are not
produced by irradiation and are probably of nucleosynthetic origin
(Fahey et al. 1987a).

The recent measurement of $^{10}$Be initial value in the hibonite
HAL makes our hypothesis amenable to confrontation with
experimental data. Marhas and Goswami (2003) have measured
$^{10}$Be/$^9$Be of (8.0 $\pm$ 3.3) $\times$ 10$^{-4}$ in HAL.
Using $^{26}$Al/$^{27}$Al = (5.4 $\pm$ 1.3) $\times$ 10$^{-8}$
(Fahey et al.~1987b), we get $\frac{^{10}\rm Be}{^{9} \rm
Be}/\frac{^{26} \rm Al}{^{27} \rm Al}$ = (8.1 $\pm$ 4) $\times$
10$^{3}$ for HAL. For gradual flares ($p=2.7$ and $^3$He/$^1$H =
0), we obtain $\frac{^{10}\rm Be}{^{9} \rm Be}/\frac{^{26} \rm
Al}{^{27} \rm Al}$ = 4.7 $\times$ 10$^{3}$. The agreement within
errors between the measured and the calculated ratio of ratios,
$\frac{^{10}\rm Be}{^{9} \rm Be}/\frac{^{26} \rm Al}{^{27} \rm
Al}$, indicates that our model can successfully account for the
decoupling between $^{26}$Al and $^{10}$Be measured by Marhas,
Goswami and Davis~(2002). Our model predicts that the
$^7$Be/$^{10}$Be ratio should be lower than normal in hibonites
and FUN inclusions.  For the sake of simplicity in the actual
calculations, we have assumed the same scaling of particle fluxes
from X-ray observations for both gradual and impulsive events, but
they could well be different.  In other words, one should not
deduce from Figure \ref{fig-all3} that the {\it absolute}
$^{10}$Be abundance is expected to be higher in gradual events
compared to impulsive events, or that the $^{10}$Be/$^{9}$Be
should be higher in FUN inclusions than in normal inclusions.

\subsection{The Astrophysical Environment}
\label{sec-context}

Irradiation in the reconnection ring produces the relatively high
energetic particles fluxes that characterize our model. It is in
this region that the magnetic fields generated by the young star
and the accretion disk reconnect and accelerate protons and helium
nuclei to energies of order of 10\,MeV or more. The reconnection
ring is confined by magnetic fields that prevent the ionized
particles from escaping. The short irradiation times used in the
model correspond to the residence times of proto-CAIs calculated
for the reconnection ring (Shu et al.~2001), and are in harmony
with the time scales observed for jets from active young stellar
objects. The absolute proton flux is obtained from the observed
X-ray luminosities of young stellar objects using a normalization
based on measurements of X-rays and particle fluxes for the
contemporary active Sun. The X-ray luminosity of active young
stellar objects estimated by Lee et al.~(1998), and used by
Gounelle et al.~(2001) as well as in this paper, is supported by
recent measurements of the X-rays emitted by young stellar objects
in the Orion Nebula Cluster (ONC) by the {\it Chandra} satellite
(Feigelson, Garmire, \& Pravdo 2002; Wolk et al.~2005). The proton
flux used in our irradiation calculations for the reconnection
ring (cited above as $F_{\rm p} (E \geq 10 {\rm MeV}) \sim 1.9
\times 10^{10}$ cm$^2$~s$^{-1}$), corresponds to an X-ray
luminosity of $\sim 5 \times 10^{30}$\,erg s$^{-1}$ (Lee et al.
1998; Gounelle et al. 2001). This is in line with the median value
for flares observed by Wolk et al. (2005), $L_X \sim 10^{31}$\,erg
s$^{-1}$ in an unprecedented Ms exposure by {\it Chandra} of the
ONC. These flares occur on an average repetition time of several
days. Much more luminous flares are observed on longer time scales
of the order of one year. Of course these observations of flares
in the ONC refer mainly to T Tauri stars with ages $\sim 1-2$\,Ma,
whereas much of the irradiation most likely occurs earlier during
a much more active stage of accretion. Following the discussion of
Lee et al (1998), the flares in that epoch were probably even more
powerful.

The irradiation model developed by Goswami, Marhas and Sahjipal
(1998), and extended by Marhas, Goswami and Davis (2002), does not
feature a favorable environment for the high particle fluxes
needed to explain the Be isotopes. These authors choose the
asteroid belt, 2-3\,AU distant from the young Sun, as the place to
irradiate meteoritic precursors by protons and alpha particles.
Because they consider the rare hibonites that contain
irradiation-produced $^{10}$Be and no $^{26}$Al to be the first
solar system solids to form (Sahjipal and Goswami 1998),
irradiation in their model takes place well before the accretion
disk was dissipated. Because a 10\,MeV proton is stopped when it
has encountered 0.2 gr\,cm$^{-2}$ of matter (Reedy 1990), the flux
at asteroidal distances is expected to be well below the level
needed to form $^7$Be and other short-lived radionuclides,
especially when the inverse-square dilution of the stellar
energetic particles is taken into account. The irradiation model
developed by Leya, Halliday and Wieler (2003) adopts the x-wind
model as a general framework for irradiation close to the young
Sun, followed by wind-borne delivery of the irradiated materials
to the asteroid belt. However, Leya et al. (2003) do not adopt a
high energy particle flux or an irradiation time from observations
or first principles, but calculate yields {\it relative} to the
$^{10}$Be/$^9$Be ratio.

\subsection{The Stellar Origin of $^{60}$Fe}

Neutron-rich $^{60}$Fe is almost impossible to synthesize in the internal
irradiation scenario because of the low abundance of appropriate
(neutron-rich) projectiles and targets (Lee et al.~1998). However,
it is well known that Type-II supernovae can produce large amounts
of $^{60}$Fe (Timmes et al.~1995; Rauscher et al.~2002). Using
updated nuclear reaction yields and improved stellar physics,
Rauscher et al (2002) estimate that Type-II supernovae can
synthesize between 2.4 $\times$ $10^{-5}$ and $16 \times$
10$^{-5}$ M$_\sun$ of $^{60}$Fe, depending on the mass of the
progenitor, somewhat higher than Timmes et al.~(1995),
0.29-10.5 $\times$ 10$^{-5}$ M$_\sun$.

The initial solar system value of $^{60}$Fe/$^{56}$Fe is
relatively well constrained since it has now been identified in
the silicate portion of chondrules that were presumably formed
early in the history of the solar system (Tachibana \&  Huss
2003). Tachibana et al.~(2005) estimate that the initial
$^{60}$Fe/$^{56}$Fe was 5-10 $\times$ 10$^{-7}$. Smaller previous
estimates were inferred by Shukolyukov \& Lugmair (1996),
$^{60}$Fe/$^{56}$Fe = 6 $\times$ 10$^{-8}$, from $^{60}$Fe
identified in planetary differentiates that formed relatively
late. The Tachibana et al.~(2005) value is compatible with the
initial $^{60}$Fe/$^{56}$Fe = 9.2 $\times$ 10$^{-6}$ estimated by
Mostefaoui et al.~(2005) from measurements in Semarkona troilite.
Using a solar system abundance of $^{56}$Fe, $^{56}$Fe/$^1$H =
$3.16 \times$ 10$^{-5}$ (Lodders 2003), the initial $^{60}$Fe
content of the solar system lies between 0.95 $\times$ 10$^{-9}$
M$_\sun$ and 1.9 $\times$ 10$^{-9}$ M$_\sun$.

The updated $^{60}$Fe yields for Type-II supernovae and the new
initial $^{60}$Fe content of the solar system have important
implications for the origin of $^{60}$Fe and other short-lived
radionuclides. They indicate that a supernova can deliver enough
$^{60}$Fe to the solar system's progenitor molecular cloud core to
account for the measured value. Other short-lived radionuclides
are not equally easily co-delivered at their initial abundances.
The contribution of a supernova, or any star, to the inventory of
short-lived radionuclides in the early solar system can be
characterized by a number of factors: the stellar yields, the mass
fraction $f$ of ejecta that is injected in the protosolar
accretion disk, and the time interval $\Delta$ between
nucleosynthesis and delivery. For example, consider the model
SN25P of Rauscher et al. (2002) that produces 1.6 $\times 10^{-4}$
M$_\sun$ of $^{60}$Fe by adopting an injection fraction $f$~=~ 1.4
$\times$ 10$^{-4}$ (2.8 $\times$ 10$^{-4}$) and a time interval
$\Delta$ = 7 Ma.  By construction, such a supernova injects 0.95
$\times$ 10$^{-9}$ M$_\sun$ (1.9 $\times$ 10$^{-9}$ M$_\sun$) of
$^{60}$Fe in the protosolar accretion disk, in line with the
observed value. Such a supernova would also deliver
$^{26}$Al, $^{36}$Cl and $^{41}$Ca at levels well below the
estimated abundances for the early solar system.

No stellar model yet published can inject into a model protosolar
accretion disk the whole inventory of short-lived radionuclides
($^{26}$Al, $^{36}$Cl, $^{41}$Ca, $^{60}$Fe). In his study of a
specific 25\,M$_{\sun}$ supernova model that delivers $^{26}$Al,
$^{41}$Ca and $^{60}$Fe at the correct abundance, Meyer (2005)
fails to deliver $^{36}$Cl by two orders of magnitude.  Moreover,
even the highest-mass stars require at least two million years of
normal stellar evolution before they supernova, and low-mass
molecular cloud-cores that form sunlike stars in Orion-like
environments (e.g., Hester et al. 2004) have disappeared before
this time.  Similarly, although specific nucleosynthetic models of
AGB stars ($M = 1.5$ M$_{\sun}$ and $Z = Z_{\sun}$/6) have the
potential to co-deliver $^{26}$Al, $^{41}$Ca and $^{60}$Fe to the
solar system, they fail to give $^{36}$Cl by two orders of
magnitude and do not produce any $^{53}$Mn (Gallino et al. 2004).
Moreover, the association of a molecular cloud core and an AGB
star is an improbable astrophysical event (e.g., Kastner and Myers
1994).  Contrary to what has been often argued (Zinner 2003;
Hester et al. 2004), the presence of $^{60}$Fe at a high level in
the nascent solar system does not necessarily imply that $^{26}$Al
and other short-lived radionuclides also have a stellar origin. A
wide range of supernovae, distant in space and time from the early
solar system, can deliver high levels of $^{60}$Fe without
producing significant amounts of other short-lived radionuclides.

\section{Conclusions}

The discovery of $^7$Be (half-life 53 days) has important
implications for early solar-system processes. First it definitely
establishes that irradiation took place at an early stage in its
formation. Second, it requires high accelerated particle fluxes,
F$_p$ (E $\geq$ 10 MeV) $\sim$ 2 $\times$ 10$^{10}$
cm$^2$~s$^{-1}$. When the measurements of $^7$Be are combined with
those for $^{10}$Be, short irradiation times of the order of a few
years to a few tens of years are adduced. The model developed by
Lee et al.~(1998) and Gounelle et al.~(2001) satisfy these
demanding conditions for high accelerated particles fluxes and
short irradiation times. Our irradiation model produces reasonable
abundances for other short-lived radionuclides, such as $^{26}$Al,
$^{36}$Cl, $^{41}$Ca and $^{53}$Mn, as well as the Be isotopes.
The overall scenario is also consistent with the evidence for
energetic photo-irradiation of early solar system rocks from the
mass-independent fractionation observed in stable sulfur and
oxygen isotopes (Rai, Jackson, \& Thiemens 2005).

The decoupling between $^{10}$Be and $^{26}$Al observed in a few
rare CAIs can be accounted for by rare gradual events. For
$^{41}$Ca, a core-mantle chemistry is required if the initial
value is as low as $^{41}$Ca/$^{40}$Ca = 1.5 $\times$ 10$^{-8}$.
If the initial abundance of $^{41}$Ca were a factor of 25 higher,
as suggested by McKeegan et al.~(2004), its abundance could be
accounted for, within the model uncertainties, by a homogeneous
chondritic composition, rather than a core-mantle structure. As
noted earlier by Lee et al.~(1998), local irradiation models fail
to produce $^{60}$Fe at the measured early solar system level,
$^{60}$Fe/$^{56}$Fe = 5-10 $\times$ 10$^{-7}$ (Tachibana~et~al.~
2005). A distant supernova is a likely origin for $^{60}$Fe, as
previously suggested by Lee et al. (1998). We note that recent
updated models of supernova nucleosynthesis produce large amounts
of $^{60}$Fe relative to other short-lived radionuclides, implying
that it is possible for supernovae to inject $^{60}$Fe in the
solar system without co-delivering other short-lived
radionuclides.

\acknowledgments The authors had illuminating discussions with K.
D. McKeegan, M. Chaussidon, A. M. Davis, R. N. Clayton, E.D. Young
and M.J. Pellin at the workshop {\it{Meteorites and the Early
Solar System}} held at Taipei, Taiwan in December 2003. They would
like to thank V. Tatischeff and J. Duprat for fruitful
conversations, as well as an anonymous reviewer for comments that
significantly improved the manuscript . This research has received
support from the Programme National de Plan\'{e}tologie (M.G.),
NASA Grant NAG5-12167 (A.G. and F.H.S.), the US Department of
Energy, Office of Nuclear Physics, under Contract No.
W-31-109-ENG-38 (KER), and by the Theoretical Institute for
Advances Research in Astrophysics (TIARA) operated under Academia
Sinica and the National Science Council Excellence Projects
through grants number NSC
94-2752-M-007-001 and  NSC 94-2112-M-007-051 (TL, FHS, and HS).  \\
This is IARC publication number 2005-0831.

\clearpage


\clearpage

\clearpage

\clearpage

\clearpage

\clearpage

\clearpage

\begin{deluxetable}{ccccccc}

\tablewidth{0pt}

\tablecaption{Initial Values of Radionuclides Considered in the
Present Paper. \label{tab-al}}

\tablehead{ \colhead{R} & \colhead{T$_{\frac{1}{2}}$} (Ma)      &
\colhead{D} & \colhead{S} & \colhead{R/S} & \colhead{ref} \\}
\startdata
$^7$Be & 53 d & $^7$Li &$^9$Be & 6.1 $\times$ 10$^{-3}$ & a\\
$^{41}$Ca & 0.1 & $^{41}$K & $^{40}$Ca & 1.5 $\times$ 10$^{-8}$ & b\\
$^{36}$Cl & 0.3 & $^{36}$S & $^{35}$Cl &  1.6 $\times$ 10$^{-4}$& c\\
$^{26}$Al & 0.74 & $^{26}$Mg & $^{27}$Al & 4.5 $\times$ 10$^{-5}$&  d\\
$^{10}$Be & 1.5 & $^{10}$B & $^{9}$Be & 1 $\times$ 10$^{-3}$ & e\\
$^{60}$Fe & 1.5 & $^{60}$Ni & $^{56}$Fe & 5-10 $\times$ 10$^{-7}$ & f\\
$^{53}$Mn & 3.7 & $^{53}$Cr & $^{55}$Mn & 4 $\times$ 10$^{-5}$ & g\\


\enddata

\tablecomments{R, D and S stand respectively for the radionuclide of
interest, its daughter and the stable reference isotope.  The R/S
values are observed or inferred in CAIs. The original
solar system values may sometimes be higher (see text). (a)
Chaussidon, Robert \& McKeegan (2005). (b) Srinivasan et
al.~(1994). (c) Lin et al.~(2005). (d) MacPherson, Zinner \& Davis
(1995). (e) McKeegan, Chaussidon \& Robert (2000). (f) Tachibana et
al.~(2005); Mostefaoui, Lugmair \& Hoppe (2005). (g) Birck \&
All\`{e}gre (1985).}

\end{deluxetable}

\clearpage

\begin{deluxetable}{lccccccc}

\tablewidth{0pt}

\tablecaption{Variation of the Beryllium-7 Irradiation Yields with
the Spectral Parameters.}

\tablehead{\colhead{} & \multicolumn{3}{c}{Core-mantle composition}
& \colhead{}
& \multicolumn{3}{c}{Chondritic composition}\\
\cline{2-4} \cline{6-8} \\
\colhead{}&\colhead{p = 2.7}&\colhead{p = 4}&\colhead{p =
5}&\colhead{} &\colhead{p = 2.7}&\colhead{p = 4}& \colhead{p = 5}}
\startdata
$^3$He/$^1$H = 0       & 1.0 (0)  &  2.3(-1) & 9.6(-2) & &9.9(-1) &2.2(-1) &9.3(-2) \\
$^3$He/$^1$H = 0.01    & 1.1 (0)  &  2.6(-1) & 1.3(-1) & &1.0(0) &2.5(-1) &1.3(-1) \\
$^3$He/$^1$H = 0.1     & 1.5(0)   &  5.6(-1) & 4.5(-1) & &1.4(0) &5.4(-1) &4.3(-1)  \\
$^3$He/$^1$H = 0.3     & 2.4(0)   &  {\bf 1.2(0)} & 1.1(0) & &2.3(0) &1.2(0) &1.1(0)  \\
$^3$He/$^1$H = 1       & 5.7(0)   &  3.5(0)  & 3.6(0)  & &5.5(0) &3.4(0)  &3.5(0)  \\

\enddata
\tablecomments{The $^7$Be/$^9$Be ratio is normalized to its
measured meteoritic value, $^7$Be/$^9$Be  = 6.1 $\times$ 10$^{-3}$. Except
for the spectral parameters, all other parameters (irradiation
time, cosmic-ray flux, etc.) are kept the same as in Gounelle et
al.~(2001) (see section 2.1). The bold type number corresponds to
the preferred case of Gounelle et al. (2001) with $p=4$ and
$^3$He/$^1$H = 0.3. \label{tab-al}}
\end{deluxetable}

\end{document}